\documentclass[aps,prb,twocolumn,superscriptaddress,showpacs]{revtex4}
\usepackage{amsmath,amssymb,amstext,graphicx,bm,bbm}
\usepackage{psfrag}
\makeatletter

\begin{document}
\title{Two-spin relaxation of P-dimers in Silicon}

\author{Massoud Borhani}
\affiliation{Department of Physics, University at Buffalo,
 SUNY, Buffalo, NY 14260-1500, USA}

\author{Xuedong Hu}
\affiliation{Department of Physics, University at Buffalo,
 SUNY, Buffalo, NY 14260-1500, USA}

\date{\today}
\pacs{03.67.Lx,72.25.Rb,76.20.+q, 76.30.Da}

\begin{abstract}
We study two-electron singlet-triplet relaxation of donor-bound electrons in Silicon.  Hyperfine interaction of the electrons with the
phosphorus (P) nuclei, in combination with the electron-phonon interaction, lead to relaxation of the triplet states. Within the Heitler-London
and effective mass approximations, we calculate the triplet relaxation rates in the presence of an applied magnetic field. This relaxation
mechanism affects the resonance peaks in current Electron Spin Resonance (ESR) experiments on P-dimers.  Moreover, the estimated time scales for
the spin decay put an upper bound on the gate pulses needed to perform fault-tolerant two-qubit operations in donor-spin-based quantum computers
(QCs).
\end{abstract}

\maketitle

Solid state based quantum information processing is a rapidly developing field, bridging quantum information science and condensed matter
physics.  Various schemes have been proposed to use spins as quantum bits (qubit), with the most prominent examples using electron spins in
quantum dots\cite{LD} and $^{31}$P nuclear spins in Si.\cite{KaneNature} Significant and exciting experimental progress have since been made to
demonstrate coherent manipulation and measurement of spins in semiconductor nanostructures.\cite{HansonRMP, FolettiNature, MorelloNature}

It has been long demonstrated experimentally that single donor electron and nuclear spins in bulk Si have extremely long coherence and
relaxation times.\cite{FeherPR} Recent experiments on isolated donors show an electron spin decoherence time $T_2$ of $0.6$ s at $1.8$ K and 0.3
Tesla magnetic field (up to 10 s at press time).\cite{Lyondonor,LyonT1T2}  This coherence time scale could be modified for donors near an
interface, although measurements for single donors near an oxide interface have obtained a promising donor electron spin relaxation time of $T_1
\sim 6$ s at a magnetic field of 1.5 Tesla.\cite{MorelloNature}

The study of spin coherence properties of phosphorus dimers (two shallow P donors that are close to each other so that they are exchange
coupled) in Si is relatively limited compared to those for single donors, because bulk experiments are ensemble averages over many different
dimer configurations and couplings, while two-donor artificial devices are still in their infancy.  P-dimers are an important ingredient in the
spin-QC architectures that use the nuclear or electron spins of $^{31}$P in Si as qubits.\cite{KaneNature,VrijenPRA}  In these proposals
exchange interaction between two bound electrons lead to an effective coupling between the corresponding electron or nuclear spins, which is
used to implement two-qubit operations.  Therefore, understanding spin coherence of P-dimers is crucial for the study of spin based quantum
computing in Si:P.

Here we calculate singlet-triplet relaxation rates of donor-bound electrons in {\it bulk} silicon within the Heitler-London
approximation.\cite{Slater} In the absence of pure dephasing (such as that due to $^{29}$Si nuclear spins), such relaxation forms one of the
decoherence channels for two-spin states of a dimer, and would give an upper bound to two-spin coherence.  This relaxation process could also
lead to broadening of resonance peaks in an ESR measurement of P-dimers, making it an observable effect within the currently available spin
resonance technology.

\begin{figure}
\begin{center}
\includegraphics[angle=0,width=0.5\textwidth]{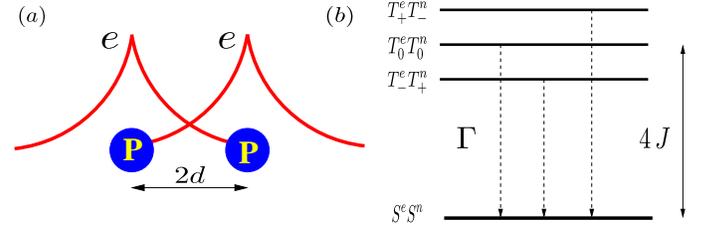}
\caption{\small (a) Schematic of a P-dimer and the corresponding (hydrogenic) electrons envelope functions. (b) Energy levels of the two
electrons  bound to a P-dimer in the quartet cluster, see Eq. (\ref{Heff}).  Mixing of the donor electron eigenstates is achieved via the
hyperfine coupling to the P nuclear spins, which causes the triplet-singlet relaxation of the electron spins via emission of acoustic phonons.
Here $\Gamma$ refers to different relaxation rates from triplet states and $4J$ is the splitting between $T^e_0$ and $S^e$.} \label{levels}
\end{center}
\end{figure}

{\it{Theoretical Model}}.  We consider two electrons which are bound to two phosphorus donors (P dimer), separated by a distance $2d$, as shown
in Fig.~\ref{levels}a.  The crystallographic axes of silicon are denoted by $x$, $y$ and $z$.  The direction of the applied magnetic field is
along $Z$ (not necessarily $z$) axis, which defines the quantization axis in the spin space. The Hamiltonian of the coupled electrons-nuclei
system is then given by $H = H_0 + V_{HF}$ where
\begin{eqnarray}
H_0 &=& g \mu_B B (\sigma^{1e}_{Z}+\sigma^{2e}_{Z}) - g_n \mu_n B (\sigma^{1n}_{Z}+\sigma^{2n}_{Z})  \nonumber \\
&& + \; J \, {\bm \sigma}^{1e} \cdot {\bm \sigma}^{2e}  \; , \\
V_{HF} &=& A_{HF} \sum_{i,j=1,2} {\bm \sigma}^{in} \cdot {\bm \sigma}^{je}  \delta ({\bm r}_j - {\bm R}_i)  ,
\end{eqnarray}
$g$ ($g_n$) is the effective g-factor of the electron (nucleus), $ \mu_B$ ($\mu_n$) is the electron (nuclear) magnetic moment, $J$ denotes the
Coulomb exchange interaction between the two electrons, and $A_{HF}=0.2$ $\mu$eV is the hyperfine coupling constant between a phosphorus nucleus
and its bound electron.\cite {KaneNature} The superscripts $e$ ($n$) refers to the electron (P nucleus), and $1$ and $2$ label the first and
the second electron (nucleus).  Notice that the total electron spin is not a good quantum number due to the hyperfine interaction $V_{HF}$. For
example, if the two nuclear spins are anti-aligned, the two electrons would experience an inhomogeneous magnetic field, which mixes the electron
singlet and triplet states.\cite{HudeSousaDasSarmaPRL}  This mixing then allows a phonon-mediated relaxation between the two-electron spin
states, and is at the core of the present study.

Using the Heitler-London approximation and averaging over the electron orbitals, we simplify the above Hamiltonian by introducing the averaged
hyperfine coupling $A$: $\langle \psi_{as} | V_{HF} |  \psi_{as} \rangle \simeq \langle \psi_{s} | V_{HF} |  \psi_{s} \rangle \simeq \frac{1}{2}
A ({\bm \sigma}^{1n} + {\bm \sigma}^{2n}) \cdot ({\bm \sigma}^{1e} + {\bm \sigma}^{2e}) $ and $\langle \psi_{as} | V_{HF} | \psi_{s} \rangle =
\langle \psi_{s} | V_{HF} | \psi_{as} \rangle \simeq \frac{1}{2} A ({\bm \sigma}^{1n} - {\bm \sigma}^{2n}) \cdot ({\bm \sigma}^{1e} - {\bm
\sigma}^{2e})$, where the symmetric and antisymmetric electronic orbital wave functions are denoted by $\psi_s$ and $\psi_{as}$, respectively.
Considering only the ground orbital state on each donor, $\psi_{a,as} = \{ \psi_L({\bm r}_1)  \psi_R({\bm r}_2) \pm  \psi_L({\bm r}_2)
\psi_R({\bm r}_1) \} / \sqrt 2$, where $\psi_L$ ($\psi_R)$ is the ground orbital wave function of the left (right) donor electron to be
specified later. We note that the first excited state of the donor bound electron is about $10$ meV above the ground state
\cite{KohnLuttinger1}, and we do not consider dimers that are too tightly coupled ($d \geq 8$ nm), so that the Heitler-London approximation is
justified for our calculation.

Within the above mentioned approximations, the $16 \times 16$ spin Hamiltonian of the two electrons and two nuclei reduces to
six clusters (one quartet, five doublets, and two singlets) that are block-diagonal. The effective Hamiltonian for the quartet cluster of the
electrons Hilbert space, in $ \{S^e S^n $, $ T_0^e T_0^n$, $T_+^e T_-^n$, $T_-^e T_+^n \}$ bases, reads
\begin{eqnarray}
H_{eff} = \left(
\begin{array}{cccc}
-3 J & 2 A & -2 A & -2 A \\
2 A & J & 2 A & 2 A \\
-2 A & 2 A & J_1 & 0 \\
-2 A & 2 A & 0 & J_2
\end{array}
\right) ,\label{Heff}\\
\nonumber\\
J_1 = J + 2 g \mu_B B + 2 g_n \mu_n B -2 A ,\label{J1}\\
J_2=  J - 2 g \mu_B B - 2 g_n \mu_n B -2 A,\label{J2}
\end{eqnarray}
where $\{ S,T_0 \} =\frac{1}{\sqrt 2} \mid\uparrow\downarrow \rangle \mp \mid\downarrow \uparrow\rangle$ and  $\{T_+, T_- \} =
\{\mid\uparrow\uparrow\rangle, \mid \downarrow\downarrow\rangle \}$ are the singlet and the triplet spin states of the electrons and the
phosphorus nuclei, as shown in Fig. \ref{levels}b.  The effective Hamiltonians for the remaining doublets are given by
\begin{eqnarray}
H^{(2)} &=& \left(
\begin{array}{cc}
-3 J & 2 A \\
2 A & J
\end{array}
\right), \{ S^eT_0^n, T_0^e S^n \}, \nonumber \\
H^{(3)} &=& \left(
\begin{array}{cc}
-3 J + 2 g_n \mu_n B & 2 A \\
2 A & J - 2 g \mu_B B
\end{array}
\right), \{ S^eT_-^n, T_-^e S^n \}, \nonumber \\
H^{(4)} &=& \left(
\begin{array}{cc}
-3 J - 2 g_n \mu_n B & 2 A \\
2 A & J +  2 g \mu_B B
\end{array}
\right), \{ S^eT_+^n, T_+^e S^n\}, \nonumber \\
 H^{(5)} &=& \left(
\begin{array}{cc}
 J + 2 g_n \mu_n B & 2 A \\
2 A & J - 2 g \mu_B B
\end{array}
\right), \{ T_0^eT_-^n, T_-^e T_0^n\}, \nonumber \\
H^{(6)} &=& \left(
\begin{array}{cc}
J - 2 g_n \mu_n B& 2 A \\
2 A & J + 2 g \mu_B B
\end{array}
\right), \{ T_0^e T_+^n, T_+^e T_0^n\}. \nonumber
\end{eqnarray}
The fully polarized states, $ \{T_+^e T_+^n\}$ and $\{ T_-^e T_-^n\} $, are decoupled from the remaining states.

In the following we neglect the Zeeman splitting of the phosphorus nuclear spins because it is about three orders of magnitude smaller than the
electron Zeeman splitting. We also assume that the averaged hyperfine coupling is much smaller than the exchange energy $A \ll J$ to keep our
perturbation treatment valid, so that level alignment is qualitatively given by Fig. \ref{levels}.b.

{\it{The electron-phonon interaction}}.  Direct relaxation of electron spin states are generally forbidden since electron-phonon interaction
conserves spin. However, as shown in Eq.~(\ref{Heff}), electron spin singlet and triplets are mixed via the hyperfine interaction in a P-dimer.
At low energies ($J \leq 1$ meV), which is our main concern here, only acoustic phonons contribute to the spin relaxation processes. Moreover,
only {\it deformation potential} coupling, consisting of a {\it dilation} part and a {\it shear} part, contribute to the relaxation processes in
Si (or Ge).
The Hamiltonian for an electron in one of the valley minima of Si is\cite{Herring,Cardona}
\begin{eqnarray}
H_{HV}^D = \Xi_d  \; {\text {Tr}}  \{ {\bm e} \} + \Xi_u (\hat {\bm k} \cdot {\bm e} \cdot \hat {\bm k}),
\end{eqnarray}
where $ {\bm e}$ is the strain tensor due to phonons and $ \hat {\bm k} $ is the unit vector along the direction of one of the [100] conduction
band minima in the reciprocal space. $\Xi_u$ is the shear deformation potential due to uniaxial strain along the [100] direction and $\Xi_d +
\Xi_u$ is the volume deformation potential\cite{Cardona}. The electron-phonon interaction for a donor electron in the ground state is then
given by\cite{Hasegawa}
\begin{eqnarray}
H_{ep}^D = \, (\Xi_d + \frac{1}{3} \Xi_u) \sum_{{\bm q}} \frac {q\; e^{i \bm {q \cdot r}}}{\sqrt {2 \rho \omega_{{\bm q}}/\hbar}} \;(a_{\bm q} +
a^{\dagger}_{-{\bm q}}), \label{Heph}
\end{eqnarray}
where $a_{\bm q}$ annihilates a phonon with momentum ${\bm q}$, $\rho=2330$ kg/m$^3$ is the silicon density, $\omega_{{\bm q}} = v_l q$ is the
energy of the longitudinal acoustic phonon, $v_l$ is its corresponding velocity and  we only consider the longitudinal phonons which have the
dominant contribution to the electron-phonon interaction. We note that in Si, $\Xi_d \sim 5.0$ eV and $\Xi_u = 8.77 $ eV.\cite{Cardona}


\begin{table}
\caption {Relaxation time $T_1$ (ms) for $T_0^eT_0^n \rightarrow S^eS^n$
(see Fig. \ref{levels}.b) as a function of $J$ (meV) and $d$ (\AA) for
{\it substitutional} donors located along [001], [011] and
[111] axes. $a_0=5.43 \, \AA$ is the lattice constant in Si,
 $a_1=a_0/\sqrt{2}$ and $a_2=\sqrt{3} a_0/4$. }

\begin{tabular}
{|c||c|c|c|c|c|c|c|c|}
\hline
$ 2 d/a_0$ & 15 & 16 & 17& 18  & 19  & 20& 21& 22 \\
\hline
$4J $ & 1.19 & 0.76 & 0.57 & 0.47 & 0.38 & 0.29 & 0.2 & 0.13  \\
\hline \hline
$T_1$ & 0.02 & 0.04 & 0.08 & 0.18 & 0.41 & 1.0 & 2.6 & 7.5 \\

\hline
\end{tabular}

\begin{tabular}
{|c||c|c|c|c|c|c|c|c|}
\hline
$ 2 d/a_1$ & 23 & 24  & 25  & 26 & 27 & 28 & 29 & 30 \\
\hline
$4J $ & 0.25 & 0.26 & 0.03 & 0.3 & 0.01 & 0.15 & 0.03 & 0.04  \\
\hline \hline
$T_1$  & 0.15 & 0.2 & 2.5 & 0.35 & 17.5 & 2 &
13 & 315  \\

\hline
\end{tabular}

\begin{tabular}
{|c||c|c|c|c|c|c|c|c|}
\hline
$ 2 d/a_2 $ & 36 & 37 & 40 & 41  & 44  & 45 & 48 & 49  \\
\hline
$4J $ & 0.26 & 0.25 &  0.34 & 0.015 & 0.06 & 0.11 & 0.006 & 0.04 \\
\hline \hline
$T_1$ & 0.08 & 0.11 & 0.18 & 5 & 2.9 & 2 & 72 &
13.5 \\

\hline
\end{tabular}
\end{table}

{\it{The relaxation rate}}.  To calculate two-spin relaxation, we need the two-electron orbital wave functions for the P-dimer.  The conduction
band of Si has six degenerate minima located close to (but not at) the edge of the first Brillouin zone.\cite{Cardona} In the effective mass
approximation, the ground state wave function of the donor bound electron is given by the homogeneous superposition of these six valleys
\begin{eqnarray}
\Psi ({\bm r}) &=& \frac{1}{\sqrt 6} \sum_{\mu =1}^6 F_{\mu}({\bm r}) \phi_{\mu}({\bm r}),
\label{wavefunction}\\
\phi_{\mu}({\bm r}) &=&  u_{\mu}({\bm r}) e^{i {\bm k}_{\mu} \cdot {\bm r}},
\end{eqnarray}
where $\mu =\{ \pm x, \pm y, \pm z \}$ is the valley index, $\phi_{\mu}({\bm r})$ are the Bloch wave functions  and $F_{\mu}({\bm r})$ are their
corresponding envelope functions. To find a simple analytical expression for the spin decay rate, we first assume Gaussian envelope functions
for the donor electrons
\begin{eqnarray}
F_{\pm z}^G ({\bm r}) = \left ( \frac{2}{\pi} \right )^{3/4}  \frac{1}{\sqrt {a^2 b}} \;
e^{-\frac{x^2 + y^2}{a^2}-\frac{z^2}{b^2}}, \label{Gaussian}
\end{eqnarray}
where $a$ and $b$ are the effective Bohr radii, and the remaining envelope functions are obtained by cyclic change of $x$, $y$ and $z$ in Eq.
(\ref{Gaussian}). Using the Fermi golden rule and Eq.~(\ref{Heph}), we calculate the relaxation rate (the one-phonon process) from  $ T_0^e$ to
$S^e$ and obtain
\begin{eqnarray}
\Gamma_{T_0 \rightarrow S} &\approx& \alpha_1 (\Xi_d + \frac{1}{3} \Xi_u)^2 \, A^2 J
(n_{q_0}+1) |I|^4 ,  \label {rate}\\
\Gamma_{T_0 \rightarrow S}^G &\approx& \alpha_2 (\Xi_d + \frac{1}{3} \Xi_u)^2 \, A^2 J
(n_{q_0}+1)\times \nonumber \\
&&\left[ e^{-d^2/b^2}+2 e^{-d^2/b^2} \right]^4,
\end{eqnarray}
where $\alpha_1 \simeq 100 \, \alpha_2= 25 / \pi^2 \rho \hbar^4 v_l^5$, $q_0 = 4 J/\hbar v_l$, $I =\langle \Psi_L ({\bm r}) |  \Psi_R ({\bm r})
\rangle$ is the overlap integral of the right and left donor wave functions, $n_q$ is the Bose-Einstein distribution of phonons, and $\Gamma^G$
is the relaxation rate for a Gaussian envelope function.  The relaxation rates $\Gamma_{T_+ \rightarrow S}$ and $\Gamma_{T_- \rightarrow S}$
have the same form as in Eq.~(\ref{rate}) but with the following replacement $J \rightarrow  (J_{1,2}+3J )/4$ [see Eqs. (\ref{J1},\ref{J2})].
Clearly, $\Gamma$ has an explicit linear dependence on the exchange energy $J$ and a quadratic dependence on the hyperfine coupling $A$.

Since in general $J \propto I^2$, one might conclude that $\Gamma$ should scale as $J^3$. However, the exchange integrals are non-trivial
functions of the distance between the donor sites $d$ and have oscillatory dependence on donor positions,\cite{HuexchangePRL} so that a simple
polynomial scaling is generally not available.

\begin{figure}
\begin{center}
\includegraphics[angle=0,width=0.5\textwidth]{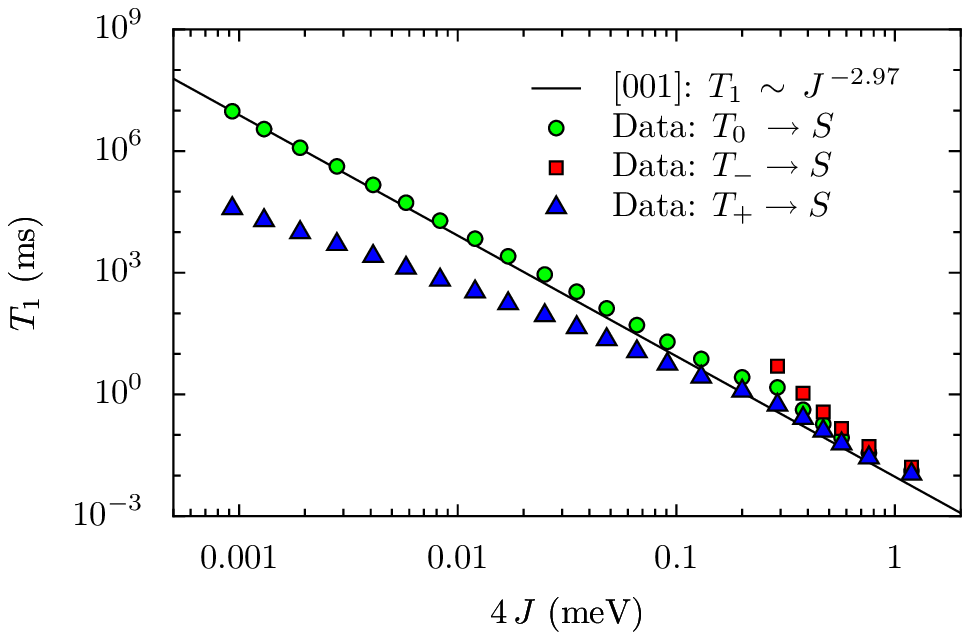}
\caption{\small (Color online) {\it Almost}-cubic dependence of the relaxation rate $\Gamma$ of the symmetric two-electron triplet state $T^e_0$
on the singlet-triplet splitting $4J$, for substitutional donors along [001] direction, see Eq.~(\ref{rate}). An external magnetic field $B = 1$
T is applied, in order to split the triplet states. Note that for $T_-^e$ state, we took only those $J$s which were larger than the Zeeman
energy.} \label{fit-T1-J}
\end{center}
\end{figure}

A more realistic envelope function $F_{\mu} ({\bm r} )$ for a donor should be hydrogenic with anisotropy.  For example, Kohn and Luttinger
proposed a variational form\cite{KohnLuttinger1}
\begin{eqnarray}
F_{\pm z} ({\bm r}) = \frac{1}{\sqrt {\pi a^2 b}} \;
e^{-\sqrt {\frac{x^2 + y^2}{a^2}+\frac{z^2}{b^2}}}. \label {hydrogenic}
\end{eqnarray}
With this envelope function there is no closed analytical form for the overlap integrals and the electron-phonon matrix elements.  We therefore
calculate the relaxation rates numerically. Table I shows our results for the relaxation rate of the unpolarized triplet to singlet state, where
we have taken into account the oscillatory behavior of $J$ as a function of the donors separation $2d$. The relaxation times change from tens of
$\mu$s to hundreds of ms as we vary the distance between the substitutional donors $2d$ (and their relative orientation) from $8$ to $12$ nm. We
also note that the value of the exchange, as a function of the donor sites, strongly depends on along which symmetry axes of the crystal the two
donors are aligned.\cite{HuexchangePRL}  We find an almost cubic dependence of the relaxation rate on $4 J$ for donors along [001] direction, as
shown in Fig. \ref{fit-T1-J}.  On the other hand, there is no reliable polynomial fit for the other two directions [011] and [111].
\begin{figure}
\begin{center}
\includegraphics[angle=0,width=0.5\textwidth]{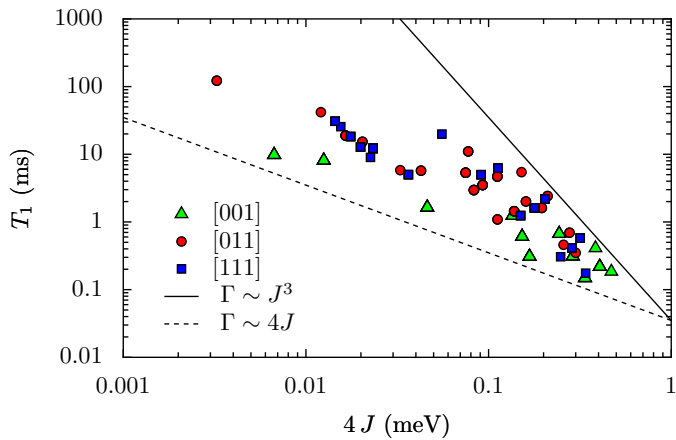}
\caption{\small (Color online) Two-spin $T_0 \rightarrow S$ relaxation times as a function of the exchange splitting for donor pairs, which are
located about $10$ nm apart from each other and along different symmetry axis.  The position of one of the donors varies within its nearest
neighboring sites, in a radius 5 Angstroms.  The solid and dotted lines show that the resulting data is within two polynomial curves of $J^3$
and $J$. } \label{variation}
\end{center}
\end{figure}

Our calculation for $J$ is based on the {\it plane wave approximation}\cite{Hudonor} of the Bloch wave function $\phi({\bm r})=u_{\bm k} ({\bm
r})e^{i{\bm k}\cdot {\bm r}}$, where we only keep the plane wave part and neglect the lattice periodic part $u_{\bm k} ({\bm r})$. The numerical
values of $J$ do not differ significantly (specifically in the Heitler-London approximation) from the more elaborate approaches which
incorporate the full Bloch structure of the Kohn-Luttinger electron wave functions\cite{Hollenberg,Hudonor}. Therefore, our final results for
the relaxation times $T_1$ give the right order of magnitude due to the linear dependence of the rates on $J$. We stress  that our calculations
are valid in the regime of relatively large exchange energies $A \ll J$. In other words, although $J$ is oscillatory as a function of $d$, it
remains always finite (over a large range of values) and larger than the hyperfine coupling in [100] direction. In contrast, for donors located
apart along [110] or [111] axes, there are certain values of $d$ which result in (almost) zero exchange energy\cite{HuexchangePRL} where our
perturbation scheme breaks down.

In reality, it is  challenging to implant donors at specific lattice sites.  Figure \ref{variation} shows the two-spin relaxation time,
calculated with Eq.~(\ref{rate}), as a function of exchange splitting $J$, for donor pairs which are about $10$ nm apart (along different
symmetry axis), with the position of one of the donors varying within a radius of $5$ Angstrom.  Our results show that even with this small
fluctuation in the donor location, there could be a three-orders-of-magnitude variation in both exchange splitting and triplet-singlet
relaxation rate, ranging from 0.1 ms to 100 ms.

The main result of our calculations is that even in purified samples (with no $^{29}$silicon isotopes), the two-electron states can relax via
the combined effect of the hyperfine interaction between phosphorus nuclei and the electrons, and the electron-phonon interaction.\cite{Siggi1}
Furthermore, this relaxation mechanism is strongly dependent on the donor separation and their orientation along different symmetry axes of the
crystal. Our results provide an upper limit to the duration of the two-qubit gates in the Kane architecture,\cite{KaneNature} where the two
phosphorus nuclei are entangled indirectly via the bound electrons through the hyperfine and exchange interaction. Due to the strong dependence
of the relaxation time of the electronic triplet states on the location (and orientation) of the donors, the physical realization of donor-based
schemes requires a careful control over the donor positions.

The data presented in Table I and Figs. (\ref{fit-T1-J},\ref{variation}), as well as the expressions for the relaxation rate $\Gamma$, allow the
identification of optimal separations and orientations for substitutional donors which mitigate this relaxation mechanism, yet still permit
sufficiently fast two-qubit operations. The key is that the relaxation rate is proportional to $J\,|I|^4$, while the speed of two-qubit gates
depends on $J$ linearly, whether for electron or nuclear spin qubits. The additional dependence on inter-donor overlap means that at larger
inter-donor separations, or for those pair-positions that have reduced overlap due to valley interference, a larger ratio of $J/\Gamma$ should
be available, so that fault-tolerant two-qubit gates are possible.  For nuclear spin qubits more specifically, we note that the use of the
electron spins as intermediaries for two-qubit gates\cite{KaneNature} could have significant impact on gate fidelity.  While isolated P nuclear
spins are known to have outstanding coherence properties,\cite{FeherPR} the electron spin relaxation mechanism studied here could put a much
stronger constraint on the gate speed, making the electron spin coherence properties the ultimate gauge to determine the feasibility of nuclear
spin qubits for quantum information processing.

In the electron spin spectroscopy for P dimers, transitions between the electron spin triplet states are detected, while singlet states are not
directly involved. The finite relaxation calculated here from triplet states to the singlet state presents a leakage for the triplet
populations, thus should lead to a broadening of the ESR signal for the dimers.

We note that the effects of the spin-orbit interaction can be safely neglected due to the small spin-orbit coupling in silicon.\cite{Abrahams}
However, for shallow donors located close to an interface, the interface roughness could potentially lead to a sufficiently strong electric
field near the bound electrons, which would in turn induce a non-negligible extrinsic spin-orbit interaction. Due to the lack of a comprehensive
model of the interface and its effect on the spin of the electrons, we did not cover this issue in this work and instead focused on bulk
properties.

In conclusion, we have calculated the relaxation rates of the electronic triplet states to the singlet state in P-dimers.  These rates have
nontrivial dependence on the singlet-triplet exchange splitting, and are strongly anisotropic. This relaxation mechanism can be studied in
P-dimers spin spectroscopy experiments by measuring the linewidth of the ESR peaks.  Finally, our results can be easily extended to similar
materials (like Ge) with different valley degeneracies.

We thank financial support from NSA/LPS through US ARO.


\end{document}